\documentclass[twocolumn,twoside,slac_two]{revtex4}

\usepackage{graphicx}
\usepackage{fancyhdr}
\def\lQ{\Lambda_{\rm QCD}}

\def\als{\alpha_{\rm s}} 

\def\siml{{\ \lower-1.2pt\vbox{\hbox{\rlap{$<$}\lower6pt\vbox{\hbox{$\sim$}}}}\ }}
\def\MS{\overline{\rm MS}}
\newcommand{\be}{\begin{equation}}
\newcommand{\ee}{\end{equation}}
\newcommand{\bea}{\begin{eqnarray}}
\newcommand{\eea}{\end{eqnarray}}

\def\simg{{\ \lower-1.2pt\vbox{\hbox{\rlap{$>$}\lower6pt\vbox{\hbox{$\sim$}}}}\ }}

\begin{document}

\title{Extraction of $\alpha_s$ and $m_Q$ from Onia}

%

\author{N. Brambilla}
\affiliation{Dipartimento di Fisica dell' Universit\'a di 
Milano and INFN, via Celoria 16,
20133 Milano, Italy}

\begin{abstract}
We briefly review how precise determinations of the strong coupling constant 
and of the heavy quark masses may be obtained from heavy quarkonium.
Such determinations are competitive with   heavy quark masses extraction 
from other systems and give an accurate value for  
the strong coupling constant at a relatively   low energy scale.
In particular we report about a recent determination of  $\als$ from 
$\Gamma(\Upsilon(1S) \to \gamma\, X)/\Gamma(\Upsilon(1S) \to X)$
with CLEO data which includes color octet contributions and avoids
model dependence in the extraction. The obtained value is $\als (M_{\Upsilon(1S)})=
0.184^{+0.015}_{-0.014}$,  which corresponds to $\als(M_Z) = 0.119^{+0.006}_{-0.005}$.
Future prospects for more precise extractions of the heavy quark masses and $\als$ are discussed.
\end{abstract}

\maketitle

\thispagestyle{fancy}


\section{QCD and the Onia}

QCD is the theory of the strong interactions: we should be able to predict 
all the properties of hadrons starting from the QCD Lagrangian which is a 
function only of the coupling constant $\als$ and of the quark masses $m$.
Therefore, from the  theoretical predictions of  physical observables and
the corresponding experimental measurements, we should be able to extract the 
values of the coupling constant $\als$ and of the quark masses $m$. However, everything 
is complicated by the fact that QCD is a strongly coupled theory in the low energy region.
At the scale $\lQ$ nonperturbative effects become dominant and $\als$ becomes large.
The nonperturbative QCD dynamics originates the confinement of quarks that in turn 
is  the reason for which the quark mass  loses its most intuitive definition.
Quarks are confined inside hadrons and thus we cannot directly measure their masses.
The mass of the quark is a parameter defined in some  renormalisation scheme at some 
renormalisation scale. 
Systems made by two heavy quarks,-quarkonia in the following-,  
 are characterized by  a quark mass scale $m_Q$  which is large, bigger than $\lQ$.
Then $\als(m_Q)$ is small and perturbative expansions may be performed at this scale.
This introduces a great simplification and hints at a factorization between 
high and low energy contributions for quarkonia. For these systems, however, things are even 
more interesting \cite{Brambilla:2004wf}.
 They are nonrelativistic systems characterized by another small parameter, 
the heavy-quark velocity $v$, and by a hierarchy of energy scales: $m_Q$ (hard),
the  relative momentum $p \sim m_Q v$ (soft),
 and the binding energy $E \sim m_Q v^2$ (ultrasoft).
For energy scales close to $\lQ$, perturbation theory breaks down  and one has to rely on nonperturbative 
methods. Regardless of this, the nonrelativistic hierarchy $m_Q \gg m_Q v \gg m_Q v^2$ 
will persist also below the $\lQ$ threshold. While the hard scale is always  larger than 
$\lQ$, different  situations may arise for the other two scales 
depending on the considered quarkonium system.
The  soft scale, proportional to the  inverse radius $r$,
may be a perturbative ($\gg \lQ$) or a nonperturbative scale ($\sim \lQ$) depending 
on the physical system. Finally, only for $t\bar{t}$ threshold states  the ultrasoft
scale  may still be perturbative. 
Heavy quark-antiquark states  probe confinement
 and nonperturbative 
physics  \cite{Brambilla:1999ja}
at different scales and are  thus an ideal and to some extent unique 
laboratory where our understanding of nonperturbative QCD, 
its interplay with perturbative QCD and the behaviour of the perturbative 
bound state series may be tested and understood in a controlled framework.
In particular in some regimes nonperturbative effects will appear  in the form 
of local or nonlocal gluon condensates and will be  suppressed in the computation 
of physical observables.
In this framework quarkonia become very appropriate systems to be used  for the study of the 
transition region from  high to low energy, for information on the QCD vacuum structure
and for precision determinations of the QCD parameters. Precisely 
this last point is the subject of this paper. In the next Sections we will discuss 
the systematic  framework offered by Non Relativistic Effective Field Theories
(NR EFT) \cite{Brambilla:2004jw}
 for the 
description of quarkonia and how one can take advantage of the accurate EFT calculations 
to make precise determinations of the QCD parameters. For some reviews of 
NR EFTs see  \cite{Brambilla:2004wf,Brambilla:2004jw,nreftrev,nreftrev2,nreftrev3}

\section{Effective Field Theories}

It is possible to 
take advantage from  the existence of a hierarchy of scales in quarkonia 
to introduce NR EFTs, which are simpler but equivalent to QCD.  
A hierarchy of EFTs may be constructed by systematically integrating out 
modes associated to  high energy scales not relevant for the quarkonium system.
Such integration  is made  in a matching procedure that 
enforces the complete equivalence between QCD and the EFT at a given 
order of the expansion in $v$ ($v^2 \sim 0.1$ for $b\bar{b}$,
 $v^2 \sim 0.3$ for $c\bar{c}$,  $v \sim 0.1$ for $t\bar{t}$).
The EFT  realizes a factorization at the Lagrangian level between 
the high energy contributions carried by matching coefficients and 
 the low energy contributions carried by the dynamical degrees of freedom.
The  Poincar\'e symmetry remains  intact in a nonlinear realization at the level of the NR EFT
imposing exact relations among the 
EFT matching coefficients \cite{poincare0,Manohar:1997qy}.

By integrating out the hard modes one  obtains Nonrelativistic QCD
 \cite{Caswell:1985ui,Bodwin:1994jh,Manohar:1997qy}.
NRQCD is making explicit at the Lagrangian level the expansions in $mv/m$ and $mv^2/m$.
It is is similar to HQET, but with a different power counting.
It also accounts for contact interactions 
between quarks and antiquark pairs (e.g. in decay processes) and hence has a wider 
set of operators.
 In NRQCD soft 
and ultrasoft scales are left dynamical and  their mixing  may complicate 
 calculations,  power counting and the consideration of the nonperturbative effects. 
In the last few years the problem of systematically treating the remaining
dynamical scales in an EFT framework has been addressed  by several groups
\cite{group,Pineda:1997bj,Brambilla:1999xf}
and has now reached a good level of understanding.
Therefore one can go down one step further and integrate out also the soft scale 
in a matching procedure to the lowest energy and simplest EFT that can be 
introduced for quarkonia,
where only ultrasoft degrees of freedom remain dynamical. 
We call such 
EFT potential NonRelativistic QCD (pNRQCD)   
\cite{Pineda:1997bj,Brambilla:1999xf} (an alternative EFT is in \cite{Luke:1999kz}).
pNRQCD is making explicit at the Lagrangian level the expansion in $mv^2/mv$.
This EFT  is close to a Schr\"odinger-like description of the bound
state and hence as simple. The bulk of the interaction
is carried by potential-like terms, but non-potential interactions,
associated with the propagation of low-energy degrees of freedom 
($Q\bar{Q}$ colour singlets, $Q\bar{Q}$  colour  octets and low energy gluons),
are generally present. They 
start  to contribute at  NLO 
in the multipole expansion of the gluon fields and are 
typically related to nonperturbative effects
\cite{Brambilla:1999xf,Brambilla:2000ch}.

In this EFT frame, 
it is important to establish when $\lQ$ sets in, i.e. when we have to 
resort to non-perturbative methods.
For low-lying resonances, it is reasonable  to assume 
$mv^2 \simg \lQ$. The system is weakly coupled and we may rely on perturbation theory,
for instance, to calculate the potential. In this case, we deal 
with weak coupling pNRQCD.
The theoretical challenge here is 
performing higher-order calculations and the goal is precision physics. This is the case 
that we will consider in this paper.

\subsection{The QCD potential and the Static Energy}

The masses may be extracted from a calculation of the 
energy levels and to obtain the energy levels we need the potential.
The $Q\bar{Q}$ potential is a Wilson  coefficient of pNRQCD 
\cite{Brambilla:1999qa}  obtained by integrating out 
all degrees of freedom but the ultrasoft ones.
It is  given by a series of contributions in an expansion in the 
inverse of the mass of the quark.
If the quarkonium system is small, the soft scale is perturbative and the 
potentials can be entirely calculated in perturbation theory 
\cite{Brambilla:2004jw}.
As matching coefficients the  potentials  undergo renormalization, 
develop a scale dependence and satisfy renormalization
group equations, which eventually allow to resum potentially large logarithms
\cite{Pineda:2001ra}.
The static singlet potential (the contribution at zero order in the mass 
expansion)  is known at three loops apart from the 
constant term
\cite{Schroder:1998vy,Brambilla:1999qa,Kniehl:2002br}. The first log related to ultrasoft effects arises at three 
loops. Such logarithm  contribution at N$^3$LO 
and the single logarithm contribution at N$^4$LO may be extracted respectively 
from a one-loop and two-loop  calculation in the EFT and have been calculated 
in \cite{Brambilla:1999qa,Brambilla:2006wp}.
The static energy 
given by the sum of a constant, the static potential and the ultrasoft corrections
is free from renormalon ambiguities. By comparing it
(at the NNLL) with lattice 
calculations one sees that the QCD perturbative series converges very nicely 
to and agrees with 
the lattice result 
\cite{Necco:2001gh}
in the short range and that no nonperturbative
linear (``stringy'') contribution to the static potential exist
\cite{Pineda:2002se,Brambilla:2004jw}. 
This is an example of how precise calculations may be performed  in this framework.
Once the renormalon contribution has been cancelled, in this case  between 
the static potential and the pole mass \cite{Hoang:1998nz,Beneke:1994sw,Brambilla:1999xf}, 
we are left with a well behaved perturbative 
series and we can unambiguosly define power corrections. It is possible to 
make predictions of physical quantities (in this case the $Q\bar{Q}$ 
static energy) at high order in the perturbative 
expansion and  with  a small error (including nonperturbative corrections 
which are suppressed in the power counting)  and to make a connection with the lattice 
results. It is remarkable that the dependence on the lattice spacing can be predicted in 
perturbation theory.

\subsection{The QCD perturbative series of the $Q\bar{Q}$
 energies and the nonperturbative contributions}
In weak coupling pNRQCD  the soft scale is perturbative  and  the potentials
are purely perturbative objects. Nonperturbative effects enter 
energy levels and decay calculations in the form of local or nonlocal 
electric and magnetic condensates 
\cite{Brambilla:1999xj,Kniehl:1999ud}.  We  still lack a precise 
and systematic knowledge  of such nonperturbative purely glue 
dependent objects. It would be important to have for them 
lattice determinations or data extraction (see e.g. \cite{Brambilla:2001xy})
The leading electric and magnetic nonlocal correlators may be related 
to the gluelump masses \cite{Brambilla:1999xf}
and to some existing lattice (quenched) determinations 
\cite{Bali:2003jq,Brambilla:2004jw}.  

However, since the nonperturbative contributions  are suppressed in the power 
counting it is possible to obtain good determinations of the masses of the
lowest quarkonium resonances  with purely perturbative calculations
in the cases in which the perturbative series is  convergent 
(after that the appropriate subtractions of renormalons have been
performed) and  large logarithms 
are resummed.  In this framework  power corrections are unambiguously defined.

\section{$m_c$ and $m_b$   extraction}

The lowest  heavy quarkonium states are   suitable systems to obtain a precise 
determination of the mass of the heavy quarks $b$ and $c$.
Perturbative determinations of the 
$\Upsilon(1S)$ and $J/\psi$ masses have been used to extract the $b$ and $c$
masses. 
These determinations  are competitive with those coming from different systems 
and different approaches (for the $b$ mass see e.g. \cite{ElKhadra:2002wp}).

Determinations of the quark masses from the perturbative calculation of $\Upsilon$
and $J/\psi$ $1S$ masses differ 
for the order of the perturbative calculation considered, for the order of the resummation of the logarithms in 
$v$   and the way in which nonperturbative corrections 
are  taken into account.  Higher order terms 
and the  residual scale dependence of the result give the theoretical error on the 
mass. The main uncertainty in these determinations
comes from nonperturbative contributions (local and nonlocal 
gluon condensates) together with possible effects due to
subleading  renormalons. 
We report some example of such determinations in Tab. 1.

\begin{table}[h]
\addtolength{\arraycolsep}{0.2cm}
\begin{center}
\begin{tabular}{|c|c|c|}
\hline
reference & order &  ${\overline m}_b({\overline m}_b)$ (GeV) 
\\
\hline
\cite{Pineda:2001zq} & NNNLO$^*$ & $4.210 \pm 0.090 \pm 0.025$
\\
\cite{Brambilla:2001qk} & NNLO +charm & $4.190 \pm 0.020 \pm 0.025$
\\
\cite{Eidemuller:2002wk} & NNLO& $4.24 \pm 0.10$
\\
\cite{Penin:2002zv} & NNNLO$^*$& $4.346 \pm 0.070$
\\
\cite{Lee:2003hh} & NNNLO$^*$ & $4.20 \pm 0.04$
\\
\cite{Contreras:2003zb} & NNNLO$^*$ & $4.241 \pm 0.070$ 
\\
\cite{Pineda:2006gx} & NNLL$^*$ & $4.19 \pm 0.06$ 
\\
\hline
\hline
reference  & order &  ${\overline m}_c({\overline m}_c)$ (GeV)  
\\
\hline
\cite{Brambilla:2001qk} & NNLO & $1.24 \pm 0.020$ 
\\
\cite{Eidemuller:2002wk}  & NNLO & $1.19 \pm 0.11$ 
\\
\hline
\end{tabular}
\vspace{2mm}
\caption{Different recent determinations of ${\overline m}_b({\overline m}_b)$
and ${\overline m}_c({\overline m}_c)$ in the $\MS$ scheme from the bottomonium and the 
charmonium systems. The displayed results use either a direct calculation 
of the lowest energy level in perturbation theory or non-relativistic 
sum rules.  
 The $^*$ indicates that the theoretical input is only partially 
complete at that order.
For  the detailed discussion about how the error has been 
computed see the original references, for a review see  \protect\cite{Brambilla:2004jw}. 
}
\end{center}  
\label{Tabmasses}
\end{table} 

Once the quark masses have been  obtained,
the renormalon subtraction and the 
same calculational approach have been  exploited  also to obtain
the energy levels of the lowest resonances.
In \cite{Brambilla:2000db}
a prediction of the $B_c$ mass has been obtained. The NNLO calculation 
with finite charm mass effects \cite{Brambilla:2001qk}
predicts  a mass of $6307(17)$ MeV that well matches the CDF 
measurement  \cite{Abulencia:2005usa}
and the lattice determination \cite{Allison:2004be}.
The same procedure seems to work at NNLO even for higher states
(inside the theory errors that grow) 
\cite{Brambilla:2001qk}.
Including logs resummation at NLL, it is possible to obtain a 
prediction for the mass of $\eta_b= 9421 \pm 11 ({\rm th}) ^{+9}_{-8} (\delta
\als)$ MeV  (where the second error comes from the uncertainty in $\als$) 
 and for 
the $B_c$ hyperfine separation $\Delta=65 \pm 24
^{+19}_{-16}$ MeV
\cite{Kniehl:2003ap}.
A NLO calculation reproduces in part the $1P$ fine splitting 
\cite{Brambilla:2004wu} .

\begin{figure*}[t]
\centering
\includegraphics[angle=270, width=16cm]{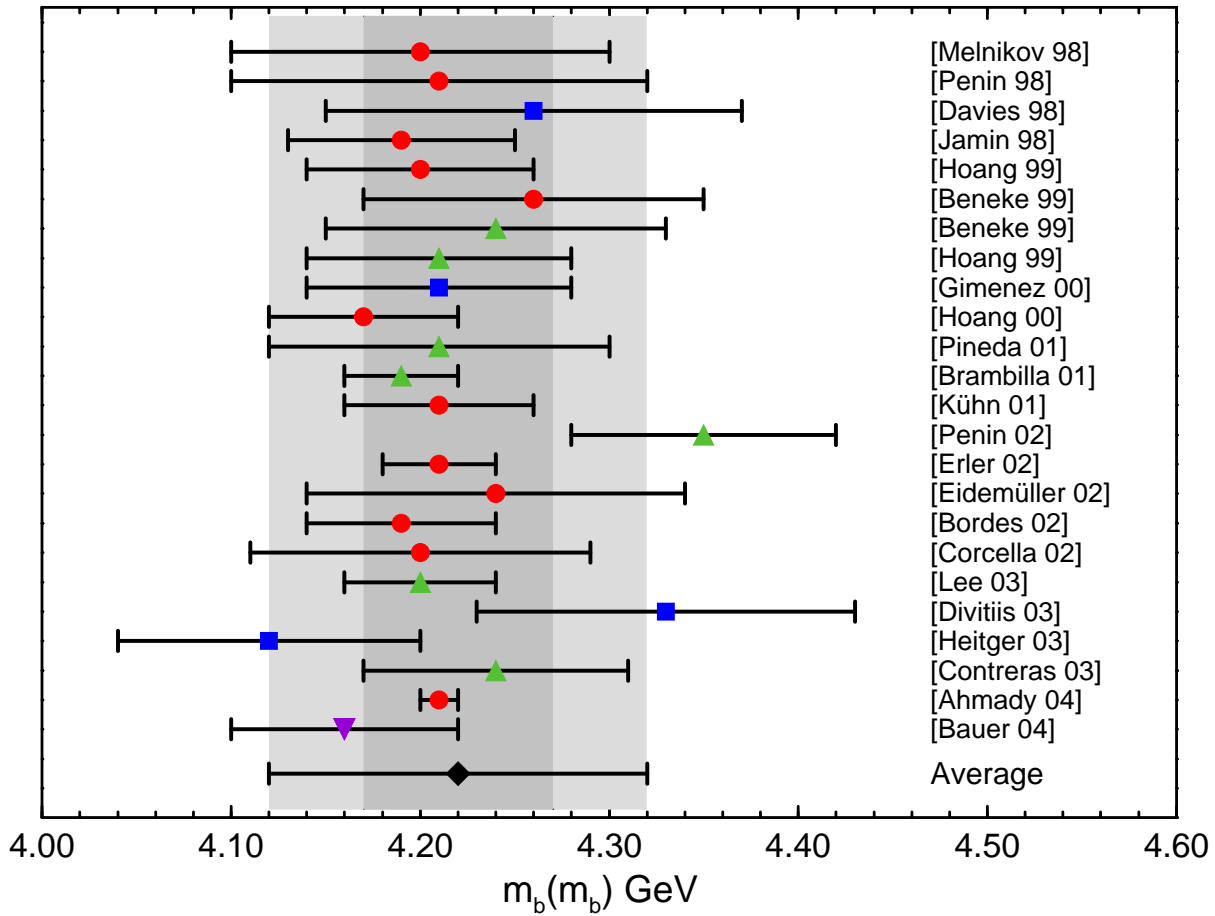}
\caption{Collection  of recent bottom quark mass
determinations. The circles represent sum rule results, the triangles
Upsilon 1S determinations, the squares lattice QCD results and the 
upside down triangle a determination from semileptonic $B$ decays. The full
diamond gives the QWG  global average for $\overline m_b(\overline m_b)$. The darker
and lighter shaded areas represent the QWG   error estimates corresponding
to a $1\sigma$ error and a range respectively.  This Table is taken from 
Chapter 6, pag. 360 of  \protect\cite{Brambilla:2004wf}. For a detailed discussion
and the explicit references to the original works
 see  \protect\cite{Brambilla:2004wf}.}\label{Tabmass1}
\end{figure*}

\begin{figure*}[t]
\centering
\includegraphics[angle=270, width=16cm]{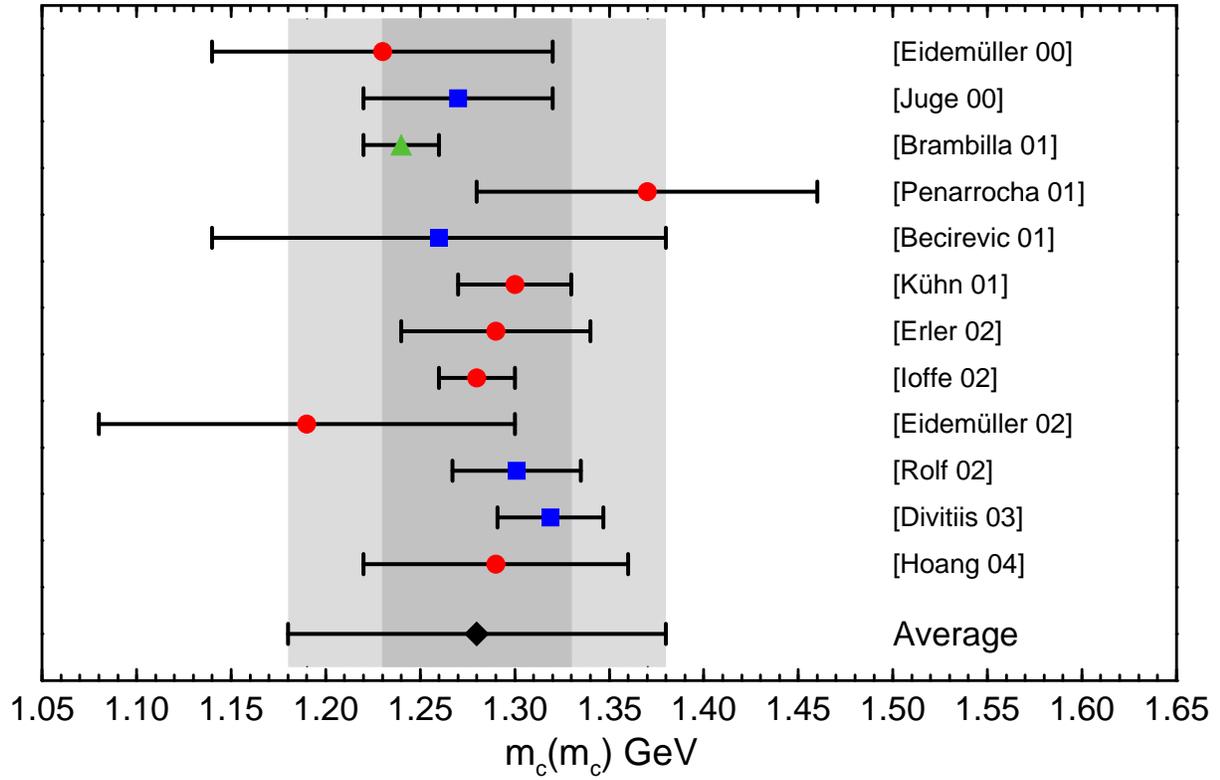}
\caption{Collection of  recent charm quark mass
determinations. The circles represent sum rule results, the  triangles
J/$\psi$ 1S determinations and the squares quenched lattice QCD results.
The full diamond gives the QWG  global average for $\overline m_c(\overline m_c)$.
The darker and lighter shaded areas represent the QWG  error estimates
corresponding to a $1\sigma$ error and a range respectively.
This Table is taken from 
Chapter 6, pag. 363 of  \protect\cite{Brambilla:2004wf}. For a detailed discussion
and the explicit references to the original works
 see  \protect\cite{Brambilla:2004wf}.}\label{Tabmass2}
\end{figure*}

A  compilation of values of the $b$ and $c$ mass has been presented by 
the Quarkonium Working Group in Chapter 6 of \cite{Brambilla:2004wf} and is reported in 
Figures \ref{Tabmass1} and \ref{Tabmass2}. 
The  mass determinations presented in such Figures  
include (relativistic and nonrelativistic) sum rule results, 
lattice QCD results, semileptonic $B$ decays as well as $\Upsilon(1S)$ 
and $J/\psi$ $1S$ determinations. One can see that the determinations 
from quarkonium are competitive with respect to determinations 
coming from other systems (heavy-light, $B$ decays).
The original works  to which  the results  in such Figures refer 
are explicitely given and discussed in \cite{Brambilla:2004wf}.
We refer to \cite{Brambilla:2004wf} also for an extended 
review of the different mass schemes, the  different heavy quark mass extractions 
approaches and the renormalon subtraction.

From these determinations the QWG reported  the following  values for  the $\MS$
masses:
\begin{eqnarray}
 \overline m_b(\overline m_b) &\,=\,& 4.22 \,\pm\, 0.05 \;{\rm GeV} \cr
 \overline m_c(\overline m_c) &\,=\,& 1.28 \,\pm\, 0.05 \;{\rm GeV} \,.
\cr
\nonumber
\end{eqnarray}
which are displayed by the darker gray area in Figures \ref{Tabmass1} and \ref{Tabmass2}
and  the following ranges:
\begin{eqnarray}
\overline m_b(\overline m_b) &\,=\,& 4.12 - 4.32 \; {\rm GeV} \cr
\overline m_c(\overline m_c) &\,=\,& 1.18 - 1.38 \;{\rm GeV} \, 
\cr
\nonumber
\end{eqnarray}
corresponding  to the lighter gray area in 
Figures \ref{Tabmass1} and \ref{Tabmass2}.
For the details of the calculation of these averages and ranges  
see \cite{Brambilla:2004wf}.

We see that the QWG values for the $b$ and $c$ mass attribute to them an  
error of $1\%$ and $4\%$ respectively. This is a  smaller error than 
the one given in the PDG  \cite{Yao:2006px}.

More recent and more accurate mass determinations 
(from lattice unquenched calculation  ${\overline m}_b({\overline m}_b) = 4.4 \pm 0.030$
GeV  \cite{Gray:2005ur};
from semileptonic  $B$ decays, 
 ${\overline m}_c({\overline m}_c) = 1.224\pm 0.017 \pm 0.054$ GeV \cite{Hoang:2005zw}; 
from low momentum sum rules
 ${\overline m}_b({\overline m}_b) = 4.164 \pm 0.025$ GeV 
${\overline m}_c({\overline m}_c) = 1.286\pm 0.013$ GeV 
\cite{Kuhn:2007vp}, and a new preliminary calculation of the mass of the b 
in the the potential subtracted scheme with unquenched lattice Fermilab action 
\cite{Freeland:2007wk})
would call for a new critical analysis and discussion of such extractions 
and   errors and an updated  mass compilation.

\subsection{$m_t$ from ttbar systems}

In \cite{Pineda:2006gx,Hoang:2001mm}
the total cross section for top quark pair production 
close to threshold in $e^+e^-$ annihilation is investigated at NNLL in the weakly coupled
EFT. The summation of the large logarithms in the ratio of the energy scales
significantly reduces the scale dependence.
Studies  like these will make  feasible a precise extractions 
of the strong coupling, the top mass 
and the top width at a future ILC. The present theoretical uncertainties for top 
mass extraction at the ILC is about 100 MeV \cite{Hoang:2006pc,Brambilla:2004jw}.

\section{$\alpha_s$ extraction from Quarkonia}

\begin{figure*}[t]
\centering
\includegraphics[angle=0, width=8cm]{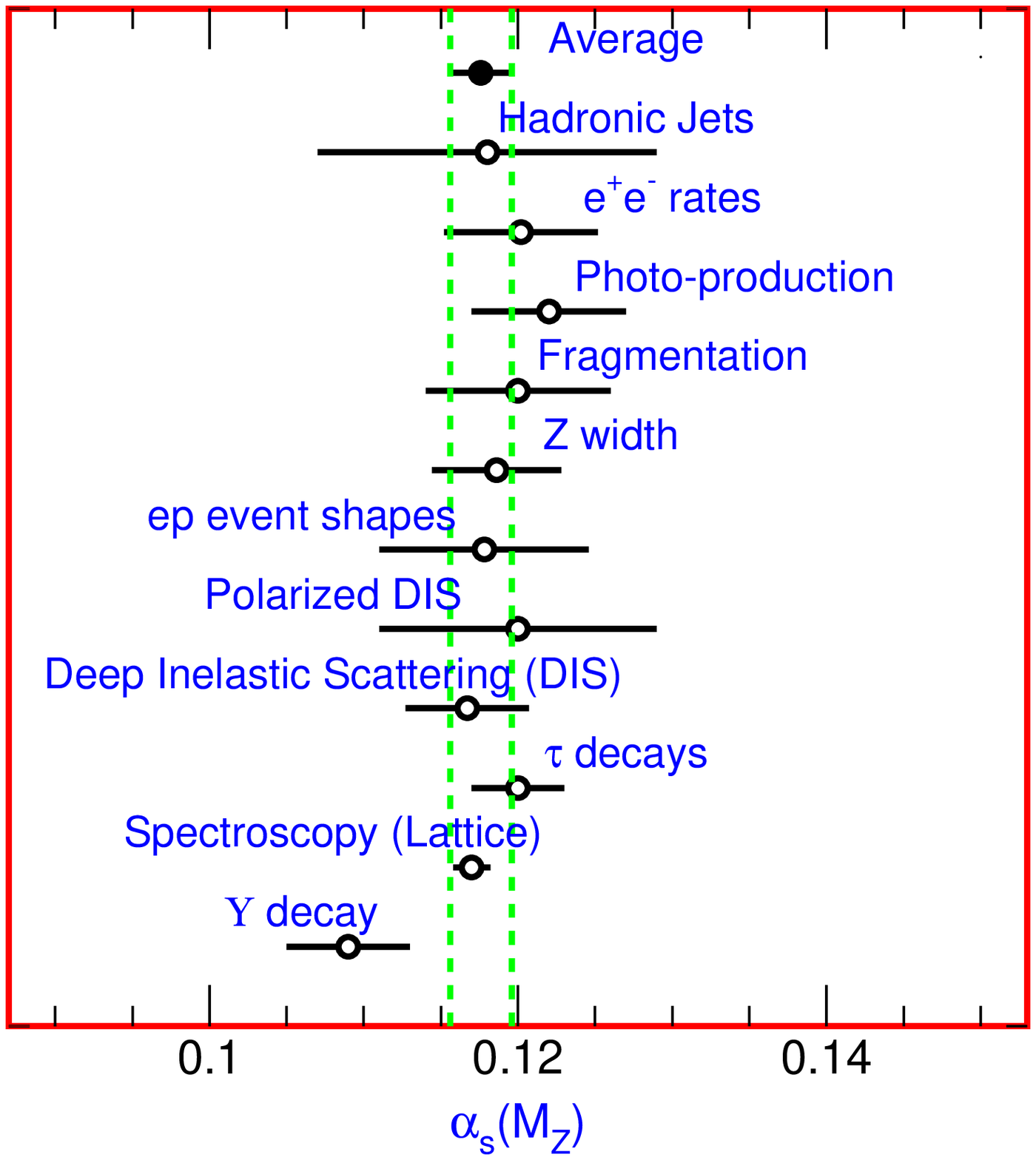}
\caption{Summary of  values of $\als(M_Z)$ from various processes, taken
  from the PDG \protect\cite{Yao:2006px}.
The value shown indicate the process and the measured value of $\als$ extrapolated to $\mu=M_Z$.
The error shown is the total error including theoretical uncertainties. The
PDG average coming from thesee measurements and quoted in the text is also shown.
Notice that the value of 
$\als$ extracted from $\Upsilon$ decays is considerably lower than all the
other determinations.} \label{figalf}
\end{figure*}

\begin{figure*}[t]
\centering
\includegraphics[angle=0, width=8cm]{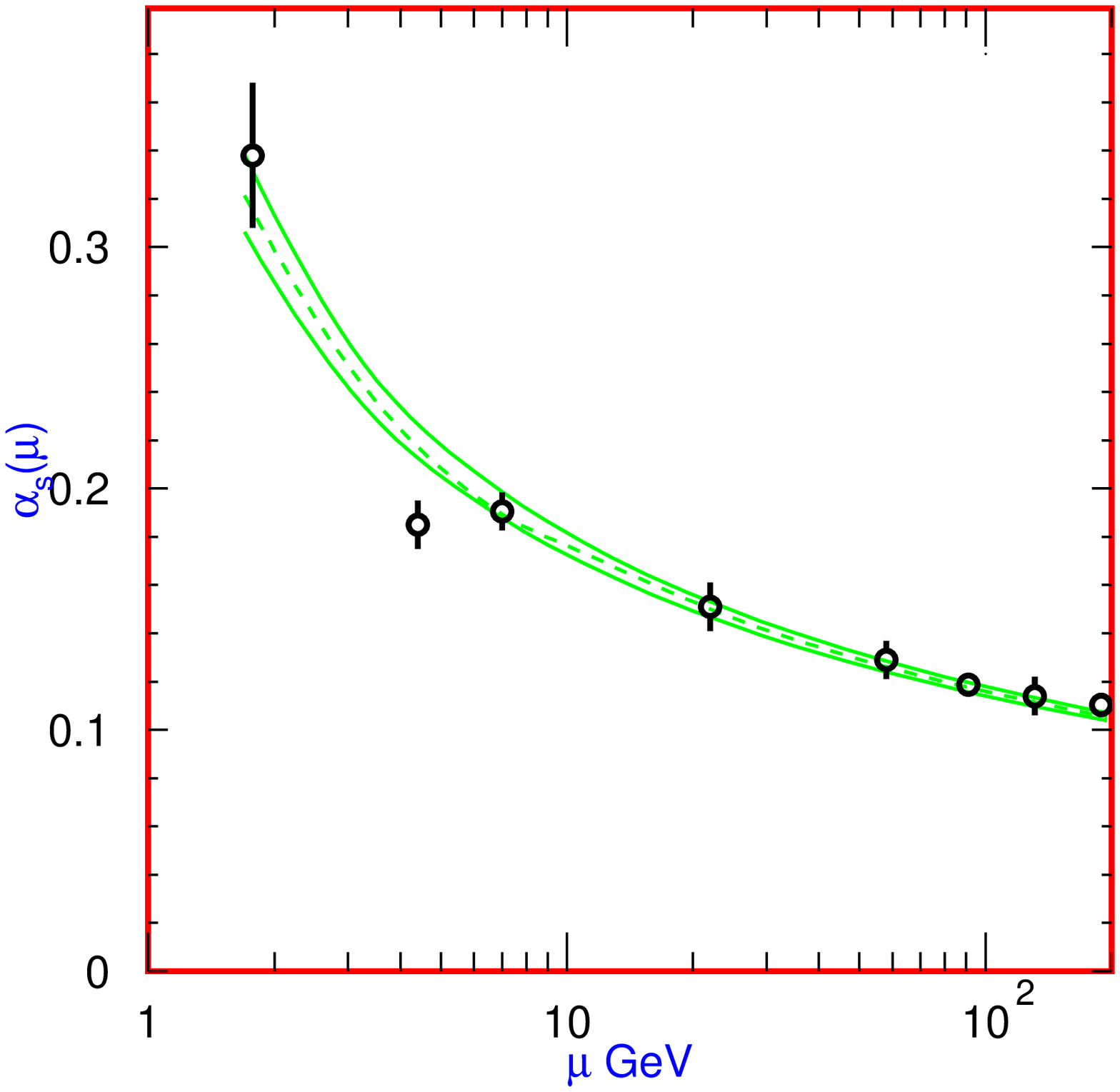}
\caption{Summary of  values of $\als(\mu)$ at the values of $\mu$ where
  they have been  measured, taken from the PDG \protect\cite{Yao:2006px}.
The line shows the central value and the $\pm 1\sigma$ limits of the PDG average.
The  data are in increasing  order of $\mu$: $\tau$ width, $\Upsilon$ decays,
  deep inelastic scattering, $e^+ e^-$ event shapes at 22 GeV from the JADE
  data, shapes at TRISTAN at 58 GeV, Z width, and $e^+ e^-$ event shapes
at 135 and 189 GeV. Notice  how the determination from the $\Upsilon$ decays 
is the only one outside the band.} \label{figalfrun}
\end{figure*}

The summary of  values of $\als(M_Z)$ from  various processes  as reported by the PDG 2006 \cite{Yao:2006px} 
is given in Fig. \ref{figalf}. We see that the value of $\als$ as determined from quarkonium 
is considerably smaller than the other determinations. The effect is seen
also in  Fig. \ref{figalfrun} where the values of 
of $\als(\mu)$ are reported at the values of $\mu$ where
  they are measured. The determination of $\als$ from  $\Upsilon $ decays is one of the
  few  ones at a relatively low energy with a relatively small error. 
 It follows from  theory calculations of ratio of hadronic and leptonic
 $\Upsilon$ decays \cite{Hinchliffe:2000yq} and use of sum rules for the $\Upsilon$ 
 system  \cite{Voloshin:1995sf,Jamin:1997rt},  the smaller error  being obtained in the first case.
Here we will report  about a determination 
 for $\als$ from the $\Upsilon$ decays
\cite{Brambilla:2007cz} that has recently solved this inconsistency.

Heavy quarkonium leptonic and non-leptonic inclusive  decay rates have historically  provided 
ways to extract  $\als$ and served as additional confirmation of the 
validity of QCD. Ratios of these quantities
  are very sensitive to $\als$  if the data are
sufficiently precise.
In particular, today  the inclusive decay widths of $J/\psi$, $\psi(2S)$ and $\Upsilon(1S)$ 
are known  with a few percent error, the ones of $\Upsilon(2S), \Upsilon(3S)$ with a 10\% 
error and most of the other inclusive decays are known 
with an error of  15-20\%. In the last few years the error on 
charmonium P-wave inclusive decays 
have been reduced to half \cite{Yao:2006px}.
On the theory side NRQCD \cite{Bodwin:1994jh}  and pNRQCD \cite{Brambilla:2002nu}
have provided powerful factorization formulas for the inclusive decays.

$S$ and $P$ wave quarkonium inclusive decays are today known in the NRQCD factorization 
up to order $v^7$ in the relativistic expansion
\cite{Brambilla:2006ph,Bodwin:2002hg,Bodwin:1994jh} and at different orders in the perturbative
expansion of the matching coefficients (see e.g. \cite{Vairo:2003gh} for a review).
 In pNRQCD the nonperturbative matrix elements 
of the four quark operators on the quarkonium states can be further decomposed in the product 
of quarkonium wave functions (or derivatives of quarkonium wave functions) in
the origin  and  
glue dependent operators, with a substantial reduction in the number of nonperturbative 
(and unknown) contributions \cite{Brambilla:2002nu}. A lattice calculation of
such  nonlocal gluonic correlators is however still missing.
 
Thanks to the EFTs  factorization between high energy contributions, calculable in 
QCD perturbation theory, and low energy nonperturbative contributions, it is possible 
to consider appropriate ratios of inclusive decays at some order of the expansion 
in $\als$  and in $v$. 
In particular, 
the ratio $\Gamma (H\rightarrow
\gamma gg)/\Gamma (H\rightarrow ggg)$ ($H$ being a quarkonium state)
appears particularly 
promising for the extraction of $\als$
\cite{Brodsky:1977du,Koller:1978qg,Voloshin:1985bd}.
since both 
the wave function at the
origin and the relativistic corrections cancel out.
However, the 
 first measurements of $J/\psi$ and $\Upsilon$ 
inclusive radiative decays  delivered a photon spectrum  not compatible
with the early QCD predictions.  Inside the EFT approach 
 it was understood that colour octet contributions, ignored in the early
calculations, become very important in the upper end-point region of the
spectrum \cite{Rothstein:1997ac}. By considering such octet contributions,
using pNRQCD to calculate them and Soft Collinear Effective Theory (SCET)  to resum end-points singularities, 
 a good description of the photon spectrum has been achieved
recently, at least for the $\Upsilon (1S)$ state
\cite{GarciaiTormo:2005ch}.  These recent theoretical advances combined 
with  new and more precise data from CLEO on $\Upsilon(1S)$ radiative decay
\cite{Besson:2005jv}, has made  the ratio $R_\gamma\equiv \Gamma(\Upsilon(1S) \to \gamma\,
  X)/\Gamma(\Upsilon(1S) \to X)$, ($X$ being hadrons)
particularly suitable 
for the $\als$ extraction at the bottom mass scale. 
For the  perturbative  calculation of the matching coefficients 
appearing in such ratio see \cite{Keung:1982jb,Bodwin:1994jh}.
Colour octet contributions also affect the ratio
$R_\gamma$ and are
parametrically of the same order of the relativistic corrections. They have so
far either been ignored \cite{Besson:2005jv} or estimated to be small
\cite{Hinchliffe:2000yq} in the available extractions of $\als$ from this
ratio. 
In \cite{Brambilla:2007cz},
recent determinations of the $\Upsilon
(1S)$ colour octet matrix elements both on the lattice \cite{Bodwin:2005gg} and
in the continuum \cite{GarciaiTormo:2004jw} have been taken into account.
This, together with the good theoretical
description \cite{GarciaiTormo:2005ch}
of the photon spectrum \cite{Besson:2005jv}, allows for a
consistent extraction of $\als (M_{\Upsilon (1S)} )$ at NLO. 
The final result obtained in \cite{Brambilla:2007cz}
is
\be 
\als (M_{\Upsilon(1S)})= 0.184^{+0.014}_{-0.013}\,,
\ee
which corresponds to 
\be 
\als (M_Z) = 0.119^{+0.006}_{-0.005}\,,
\ee 
 very close to the central value of the PDG \cite{Yao:2006px} with competitive errors. The key
ingredients to get these numbers have been the precise CLEO data
\cite{Besson:2005jv}, the use of a QCD calculation  to extrapolate the photon spectrum at low $z$, and
accurate estimates of the colour octet matrix elements, which have been possible
thanks to recent lattice and continuum calculations.
At present, the main uncertainty 
in the extraction  of $\als$ comes from  the systematic uncertainties in
${R_\gamma^{\rm exp}}$.

The impact of this  determination of $\als$, if included in the world average,
will be to increase it.

\section{Future Prospects for mass and $\als$ extractions}

The mass and $\als$ determinations from quarkonium that we have presented 
are  already competitive with the results obtained from other physical systems.

In the near future $\als(m_c)$ may be extracted  from the $R_\gamma$ 
ratio for the $J/\psi$  provided that a new measurement of the inclusive photon 
spectrum for radiative $J/\psi$ decays will be performed at BESIII
\cite{GarciaiTormo:2007qs}. In a similar way, the discovery and the
measurement of the $\eta_b$  mass with a few MeV accuracy will  provide 
 a  determination of $\als(M_Z)$  with 3 per mille error
 from the hyperfine 
separation calculated at NLL \cite{Kniehl:2003ap}.

For  an improved  determination of $\als$ from the lattice calculations of the 
quarkonium  spectrum,  we need a nonperturbative unquenched determination of
$\Lambda_{\overline{\rm MS}}$  and results on the spectrum obtained with
different  formulations of sea quarks, besides staggered quarks. Also the improvement in the lattice 
extraction of the masses would require an improved accuracy in the conversion
from the bare lattice mass to  the $\overline{\rm MS}$ mass. In particular the
two loop matching in such conversion would be needed for the Fermilab and the
NRQCD actions. A nonperturbative matching would also be desiderable.

For what concerns the  mass extraction from the $\Upsilon(1S)$ and $J/\psi$
masses in perturbation theory  at present, as it has been discussed, the
major theoretical error comes from  our ignorance of the ultrasoft
nonperturbative corrections. A lattice calculation of the
nonperturbative chromoelectric correlator together with  its  matching from lattice
to $\overline{\rm MS}$ scheme  is needed.

Further improvements in the mass determinations from  nonrelativistic sum rule
would require the full NNLL calculation; a complete NNNLO computation would
also be useful to have a better control on the 
theoretical uncertainties. For low momentum sum rules, improved determinations
of the $R$ measurements around   bottomonium and charmonium region 
would be crucial.

We conclude  noticing that, within the EFT approach and the factorization 
scheme,  precision calculations in quarkonium 
may be applied to all the  physical observables of the lowest resonances, spectra  and
decays included. To this respect it is particularly interesting 
the example of the calculation of M1 transitions  for the lowest 
quarkonia resonances.  In this case the Poincar\`e invariance of the EFT 
imposes exact relations among matching coefficients that set to zero 
the nonperturbative corrections at order $v^2$. At this order the M1
transitions  may be exactly calculated in perturbation theory
\cite{Brambilla:2005zw}.

\begin{acknowledgments}
 Support inside the  European 
Research Training Network FLAVIA{\it net} (FP6, Marie Curie Programs, Contract 
MRTN-CT-2006-035482) is 
acknowledged. I thank the organizers for the organization of this very interesting 
and nice meeting.
\end{acknowledgments}


\begin{thebibliography}{99} 



\bibitem{Brambilla:2004wf}
N.~Brambilla {\it et al.},
``Heavy quarkonium physics,''
CERN-2005-005, (CERN, Geneva, 2005)
[arXiv:hep-ph/0412158].
See also the web page of the International Quarkonium
Working Group: http://www.qwg.to.infn.it.

\bibitem{Brambilla:1999ja}
  N.~Brambilla and A.~Vairo,
  arXiv:hep-ph/9904330.

\bibitem{Brambilla:2004jw}
N.~Brambilla, A.~Pineda, J.~Soto and A.~Vairo,
Rev.\ Mod.\ Phys.\  {\bf 77}, 1423 (2005)
[arXiv:hep-ph/0410047].


\bibitem{nreftrev}
  J.~Soto,
  arXiv:nucl-th/0611055;
  A.~Vairo,
  arXiv:hep-ph/0610251;
  P.~Ruiz-Femenia and A.~Hoang,
  arXiv:hep-ph/0611291;
  A.~V.~Manohar and I.~W.~Stewart,
  Phys.\ Rev.\  D {\bf 76}, 074002 (2007)
  [arXiv:hep-ph/0605001].
  N.~Brambilla,
  arXiv:hep-ph/0012026.


\bibitem{nreftrev2}
  G.~T.~Bodwin,
  Int.\ J.\ Mod.\ Phys.\  A {\bf 21}, 785 (2006)
  [arXiv:hep-ph/0509203].



\bibitem{nreftrev3}
  G.~P.~Lepage,
  arXiv:hep-ph/0506330;
  B.~Grinstein,
  Int.\ J.\ Mod.\ Phys.\  A {\bf 15}, 461 (2000)
  [arXiv:hep-ph/9811264];
  U.~Van Kolck, L.~J.~Abu-Raddad and D.~M.~Cardamone,
  arXiv:nucl-th/0205058;
  A.~H.~Hoang,
  arXiv:hep-ph/0204299.
  I.~Z.~Rothstein,
  arXiv:hep-ph/0308266.



\bibitem{poincare0}
N.~Brambilla, D.~Gromes and A.~Vairo,
Phys.\ Lett.\ B {\bf 576}, 314 (2003);
Phys.\ Rev.\ D {\bf 64}, 076010 (2001).



\bibitem{Caswell:1985ui}
  W.~E.~Caswell and G.~P.~Lepage,
  Phys.\ Lett.\  B {\bf 167}, 437 (1986).

\bibitem{Bodwin:1994jh}
G.~T.~Bodwin, E.~Braaten and G.~P.~Lepage,
Phys.\ Rev.\ D {\bf 51}, 1125 (1995)
[Erratum-ibid.\ D {\bf 55}, 5853 (1997)]
[arXiv:hep-ph/9407339].

\bibitem{Manohar:1997qy}
  A.~V.~Manohar,
  Phys.\ Rev.\  D {\bf 56}, 230 (1997)
  [arXiv:hep-ph/9701294].



\bibitem{group}
  M.~Beneke and V.~A.~Smirnov,
  Nucl.\ Phys.\ B {\bf 522}, 321 (1998);
  B.~A.~Kniehl and A.~A.~Penin,
  Nucl.\ Phys.\ B {\bf 563}, 200 (1999);
  M.~E.~Luke and A.~V.~Manohar,
  Phys.\ Rev.\ D {\bf 55}, 4129 (1997);






\bibitem{Pineda:1997bj}
  A.~Pineda and J.~Soto,
  Nucl.\ Phys.\ Proc.\ Suppl.\  {\bf 64}, 428 (1998)
  [arXiv:hep-ph/9707481].

\bibitem{Brambilla:1999xf}
  N.~Brambilla, A.~Pineda, J.~Soto and A.~Vairo,
  Nucl.\ Phys.\ B {\bf 566}, 275 (2000) [arXiv:hep-ph/9907240].


\bibitem{Luke:1999kz}
  M.~E.~Luke, A.~V.~Manohar and I.~Z.~Rothstein,
  Phys.\ Rev.\ D {\bf 61}, 074025 (2000).
  A.~V.~Manohar and I.~W.~Stewart,
  Phys.\ Rev.\  D {\bf 62}, 074015 (2000)
  [arXiv:hep-ph/0003032].




\bibitem{Pineda:2001ra}
A.~Pineda,
Phys.\ Rev.\ D {\bf 65}, 074007 (2002);
A.~Pineda and J.~Soto,
Phys.\ Lett.\ B {\bf 495}, 323 (2000);
  A.~H.~Hoang and I.~W.~Stewart,
  Phys.\ Rev.\ D {\bf 67}, 114020 (2003);
  A.~V.~Manohar and I.~W.~Stewart,
  Phys.\ Rev.\ D {\bf 62}, 014033 (2000).


\bibitem{Brambilla:2000ch}
  N.~Brambilla,
  arXiv:hep-ph/0012211.


\bibitem{Brambilla:1999qa}
  N.~Brambilla, A.~Pineda, J.~Soto and A.~Vairo,
  Phys.\ Rev.\ D {\bf 60}, 091502 (1999).

\bibitem{Schroder:1998vy}
  Y.~Schroder,
  Phys.\ Lett.\  B {\bf 447}, 321 (1999)
  [arXiv:hep-ph/9812205];
  M.~Peter,
  Nucl.\ Phys.\  B {\bf 501}, 471 (1997)
  [arXiv:hep-ph/9702245].
  F.~A.~Chishtie and V.~Elias,
  Phys.\ Lett.\  B {\bf 521}, 434 (2001)
  [arXiv:hep-ph/0107052].



\bibitem{Brambilla:2006wp}
  N.~Brambilla, X.~Garcia i Tormo, J.~Soto and A.~Vairo,
  Phys.\ Lett.\  B {\bf 647}, 185 (2007).





\bibitem{Pineda:2002se}
  A.~Pineda,
  J.\ Phys.\ G {\bf 29}, 371 (2003).


\bibitem{Hoang:1998nz}
  A.~H.~Hoang, M.~C.~Smith, T.~Stelzer and S.~Willenbrock,
  Phys.\ Rev.\  D {\bf 59}, 114014 (1999)
  [arXiv:hep-ph/9804227].
  A.~Pineda,
  ``Heavy quarkonium and nonrelativistic effective field theories,'',
 PH. D. Thesis, Barcelona 1998.


\bibitem{Beneke:1994sw}
  M.~Beneke and V.~M.~Braun,
  Nucl.\ Phys.\  B {\bf 426}, 301 (1994)
  [arXiv:hep-ph/9402364].
  I.~I.~Y.~Bigi, M.~A.~Shifman, N.~G.~Uraltsev and A.~I.~Vainshtein,
  Phys.\ Rev.\  D {\bf 50}, 2234 (1994)
  [arXiv:hep-ph/9402360].


\bibitem{Brambilla:1999xj}
  N.~Brambilla, A.~Pineda, J.~Soto and A.~Vairo,
  Phys.\ Lett.\  B {\bf 470}, 215 (1999).

\bibitem{Kniehl:1999ud}
  B.~A.~Kniehl and A.~A.~Penin,
  Nucl.\ Phys.\  B {\bf 563}, 200 (1999)
  [arXiv:hep-ph/9907489].


\bibitem{Kniehl:2002br}
  B.~A.~Kniehl, A.~A.~Penin, V.~A.~Smirnov and M.~Steinhauser,
  Nucl.\ Phys.\  B {\bf 635}, 357 (2002)
  [arXiv:hep-ph/0203166].


\bibitem{Necco:2001gh}
  S.~Necco and R.~Sommer,
  Phys.\ Lett.\  B {\bf 523}, 135 (2001)
  [arXiv:hep-ph/0109093].
  G.~S.~Bali, K.~Schilling and A.~Wachter,
  Phys.\ Rev.\  D {\bf 56}, 2566 (1997)
  [arXiv:hep-lat/9703019].





\bibitem{Brambilla:2001xy}
  N.~Brambilla, D.~Eiras, A.~Pineda, J.~Soto and A.~Vai\-ro,
  Phys.\ Rev.\ Lett.\  {\bf 88}, 012003 (2002).



\bibitem{Bali:2003jq}
  G.~S.~Bali and A.~Pineda,
  Phys.\ Rev.\  D {\bf 69}, 094001 (2004)
  [arXiv:hep-ph/0310130].




\bibitem{Brambilla:2000db}
  N.~Brambilla and A.~Vairo,
  Phys.\ Rev.\ D {\bf 62}, 094019 (2000).




\bibitem{Brambilla:2001qk}
  N.~Brambilla, Y.~Sumino and A.~Vairo,
  Phys.\ Rev.\ D {\bf 65}, 034001 (2002);
  N.~Brambilla, Y.~Sumino and A.~Vairo,
  Phys.\ Lett.\ B {\bf 513}, 381 (2001).


\bibitem{Abulencia:2005usa}
  A.~Abulencia {\it et al.}  [CDF Collaboration],
  Phys.\ Rev.\ Lett.\  {\bf 96}, 082002 (2006).



\bibitem{Allison:2004be}
  I.~F.~Allison, C.~T.~H.~Davies, A.~Gray, A.~S.~Kronfeld, P.~B.~Mackenzie and J.~N.~Simone
                  [HPQCD Collaboration],
  Phys.\ Rev.\ Lett.\  {\bf 94}, 172001 (2005).



\bibitem{Kniehl:2003ap}
  B.~A.~Kniehl, A.~A.~Penin, A.~Pineda, V.~A.~Smirnov and M.~Steinhauser,
  Phys.\ Rev.\ Lett.\  {\bf 92}, 242001 (2004);
  [arXiv:hep-ph/0312086];
  A.~A.~Penin, A.~Pineda, V.~A.~Smirnov and M.~Steinhauser,
  Phys.\ Lett.\  B {\bf 593}, 124 (2004).


\bibitem{Brambilla:2004wu}
  N.~Brambilla and A.~Vairo,
  Phys.\ Rev.\  D {\bf 71}, 034020 (2005).


\bibitem{ElKhadra:2002wp}
  A.~X.~El-Khadra and M.~Luke,
  Ann.\ Rev.\ Nucl.\ Part.\ Sci.\  {\bf 52}, 201 (2002)
  [arXiv:hep-ph/0208114].







\bibitem{Pineda:2001zq}
  A.~Pineda,
  JHEP {\bf 0106}, 022 (2001)
  [arXiv:hep-ph\-/01\-05\-008].


\bibitem{Penin:2002zv}
  A.~A.~Penin and M.~Steinhauser,
  Phys.\ Lett.\ B {\bf 538}, 335 (2002)
  [arXiv:hep-ph/0204290].

\bibitem{Eidemuller:2002wk}
  M.~Eidem\"uller,
  Phys.\ Rev.\ D {\bf 67}, 113002 (2003)
  [arXiv:hep-ph/0207237].

\bibitem{Lee:2003hh}
  T.~Lee,
  JHEP {\bf 0310}, 044 (2003)
  [arXiv:hep-ph/0304185].

\bibitem{Contreras:2003zb}
  C.~Contreras, G.~Cvetic and P.~Gaete,
  Phys.\ Rev.\ D {\bf 70}, 034008 (2004)
  [arXiv:hep-ph/0311202].

\bibitem{Pineda:2006gx}
  A.~Pineda and A.~Signer,
  Phys.\ Rev.\ D {\bf 73}, 111501 (2006)
  [arXiv:hep-ph/0601185].



\bibitem{Yao:2006px}
 W.~M.~Yao {\it et al.}  [Particle Data Group],
 J.\ Phys.\ G {\bf 33}, 1 (2006). 

\bibitem{Hoang:2005zw}
  A.~H.~Hoang and A.~V.~Manohar,
  Phys.\ Lett.\  B {\bf 633}, 526 (2006)
  [arXiv:hep-ph/0509195].


\bibitem{Gray:2005ur}
  A.~Gray, I.~Allison, C.~T.~H.~Davies, E.~Dalgic, G.~P.~Lepage, J.~Shigemitsu and M.~Wingate,
  Phys.\ Rev.\  D {\bf 72}, 094507 (2005)
  [arXiv:hep-lat/0507013].


\bibitem{Kuhn:2007vp}
  J.~H.~Kuhn, M.~Steinhauser and C.~Sturm,
  Nucl.\ Phys.\  B {\bf 778}, 192 (2007)
  [arXiv:hep-ph/0702103].

\bibitem{Freeland:2007wk}
  E.~D.~Freeland, A.~S.~Kronfeld, J.~N.~Simone and R.~S.~Van de Water
                  [Fermilab Lattice Collaboration],
  arXiv:0710.4339 [hep-lat].


\bibitem{Hoang:2006pc}
  A.~H.~Hoang,
  PoS {\bf TOP2006}, 032 (2006)
  [arXiv:hep-ph/0604185].


\bibitem{Hoang:2001mm}
  A.~H.~Hoang, A.~V.~Manohar, I.~W.~Stewart and T.~Teubner,
  Phys.\ Rev.\  D {\bf 65}, 014014 (2002)
  [arXiv:hep-ph/0107144].




\bibitem{Brambilla:2002nu}
  N.~Brambilla, D.~Eiras, A.~Pineda, J.~Soto and A.~Vairo,
  Phys.\ Rev.\  D {\bf 67}, 034018 (2003)
  [arXiv:hep-ph/0208019]
  N.~Brambilla, A.~Pineda, J.~Soto and A.~Vairo,
  Phys.\ Lett.\  B {\bf 580}, 60 (2004)
  [arXiv:hep-ph/0307159].

\bibitem{Brambilla:2006ph}
  N.~Brambilla, E.~Mereghetti and A.~Vairo,
  JHEP {\bf 0608}, 039 (2006)
  [arXiv:hep-ph/0604190];
 N.~Brambilla, E.~Mereghetti and A.~Vairo,
``Hadronic quarkonium decays at order v**7,'' IFUM-899-FT.


\bibitem{Brodsky:1977du}
 S.~J.~Brodsky, D.~G.~Coyne, T.~A.~DeGrand and R.~R.~Horgan,
 Phys.\ Lett.\ B {\bf 73}, 203 (1978).

\bibitem{Koller:1978qg}
 K.~Koller and T.~Walsh,
 Nucl.\ Phys.\ B {\bf 140}, 449 (1978).




\bibitem{Rothstein:1997ac}
 I.~Z.~Rothstein and M.~B.~Wise,
 Phys.\ Lett.\ B {\bf 402}, 346 (1997) 
 [arXiv:hep-ph/9701404].

\bibitem{GarciaiTormo:2005ch}
 X.~Garcia i Tormo and J.~Soto,
 Phys.\ Rev.\ D {\bf 72}, 054014 (2005) 
 [arXiv:hep-ph/0507107].

\bibitem{Besson:2005jv}
 D.~Besson {\it et al.}  [CLEO Collaboration],
 Phys.\ Rev.\ D {\bf 74}, 012003 (2006) 
 [arXiv:hep-ex/0512061].


\bibitem{Keung:1982jb}
  W.~Y.~M.~Keung and I.~J.~Muzinich,
  Phys.\ Rev.\ D {\bf 27}, 1518 (1983); 
 F.~Maltoni and A.~Petrelli,
 Phys.\ Rev.\ D {\bf 59}, 074006 (1999) 
 [arXiv:hep-ph/9806455];
  A.~Petrelli, M.~Cacciari, M.~Greco, F.~Maltoni and M.~L.~Mangano,
  Nucl.\ Phys.\ B {\bf 514}, 245 (1998) 
  [arXiv:hep-ph/9707223];
 M.~Kr\"amer,
 Phys.\ Rev.\ D {\bf 60}, 111503 (1999) 
 [arXiv:hep-ph/9904416];
 P.~B.~Mackenzie and G.~P.~Lepage,
 Phys.\ Rev.\ Lett.\  {\bf 47}, 1244 (1981); 
  J.~Campbell, F.~Maltoni and F.~Tramontano,
  Phys.\ Rev.\ Lett.\  {\bf 98}, 252002 (2007)
  [arXiv:hep-ph/0703113].





\bibitem{Hinchliffe:2000yq}
 I.~Hinchliffe and A.~V.~Manohar,
 Ann.\ Rev.\ Nucl.\ Part.\ Sci.\  {\bf 50}, 643 (2000) 
 [arXiv:hep-ph/0004186].



\bibitem{Voloshin:1995sf}
  M.~B.~Voloshin,
  Int.\ J.\ Mod.\ Phys.\  A {\bf 10}, 2865 (1995)
  [arXiv:hep-ph/9502224].

\bibitem{Jamin:1997rt}
  M.~Jamin and A.~Pich,
  Nucl.\ Phys.\  B {\bf 507}, 334 (1997)
  [arXiv:hep-ph/9702276].


\bibitem{Bodwin:2005gg}
 G.~T.~Bodwin, J.~Lee and D.~K.~Sinclair,
 Phys.\ Rev.\ D {\bf 72}, 014009 (2005) 
 [arXiv:hep-lat/0503032].

\bibitem{GarciaiTormo:2004jw}
 X.~Garcia i Tormo and J.~Soto,
 Phys.\ Rev.\ D {\bf 69}, 114006 (2004) 
 [arXiv:hep-ph/0401233].






\bibitem{Brambilla:2007cz}
  N.~Brambilla, X.~Garcia i Tormo, J.~Soto and A.~Vairo,
  Phys.\ Rev.\  D {\bf 75}, 074014 (2007)
  [arXiv:hep-ph/0702079].


\bibitem{Voloshin:1985bd}
  M.~B.~Voloshin,
  Sov.\ J.\ Nucl.\ Phys.\  {\bf 40}, 662 (1984)
  [Yad.\ Fiz.\  {\bf 40}, 1039 (1984)].


\bibitem{Bodwin:2002hg}
  G.~T.~Bodwin and A.~Petrelli,
  Phys.\ Rev.\  D {\bf 66}, 094011 (2002)
  [arXiv:hep-ph/0205210].


\bibitem{Vairo:2003gh}
  A.~Vairo,
  Mod.\ Phys.\ Lett.\  A {\bf 19}, 253 (2004)
  [arXiv:hep-ph/0311303].

\bibitem{GarciaiTormo:2007qs}
  X.~Garcia i Tormo and J.~Soto,
  arXiv:hep-ph/0701030.


\bibitem{Brambilla:2005zw}
  N.~Brambilla, Y.~Jia and A.~Vairo,
  Phys.\ Rev.\  D {\bf 73}, 054005 (2006)
  [arXiv:hep-ph/0512369].



\end{thebibliography}
\end{document}